\documentclass[11pt,fleqn]{article}
\usepackage{times}
\usepackage{graphics,latexsym,psfrag}
\usepackage{amsmath}
\usepackage{amssymb}
\usepackage{graphicx} 
\usepackage{color}
\usepackage{caption} 
\usepackage{subcaption} 
\usepackage{tabularx} 
\usepackage{multirow}

\title{\bf A two-phase two-layer model for transdermal  \\
 drug delivery and percutaneous absorption }
\author{{\em Giuseppe Pontrelli\footnote{Corresponding author.}} \vspace{1mm}\\ 
Istituto per le Applicazioni del Calcolo - CNR\\
Via dei Taurini 19 -- 00185 Roma, Italy
\vspace{1mm}\\
E-mail: {\tt giuseppe.pontrelli@gmail.com}
\vspace{3mm}\\
{\em Filippo de Monte} \vspace{1mm}\\ 
Department of Industrial and Information Engineering and Economics\\
 University of L'Aquila \\
Via G. Gronchi 18 -- 67100 L'Aquila, Italy
\vspace{1mm}\\
E-mail: {\tt filippo.demonte@univaq.it}
}
\date{}
\newcommand{\be}{\begin{equation}}
\newcommand{\ee}{\end{equation}}
\newcommand{\bc}{\begin{center}}
\newcommand{\ec}{\end{center}}
\newcommand{\bdm}{\begin{displaymath}}
\newcommand{\edm}{\end{displaymath}}
\newcommand{\p}{\partial}

\newcommand{\h}{\backslash}

\newcommand{\ds}{\displaystyle}

\newcommand{\INT}{\int\limits}
\newcommand{\SUM}{\sum\limits}

\begin{document}
\maketitle

\begin{abstract}
One of the promising frontiers of bioengineering 
is the controlled release of a therapeutic
drug from a vehicle across the skin (transdermal drug delivery). 
In order to study the complete process, a two-phase mathematical model 
describing the dynamics of a substance between
two  coupled  media of different properties and dimensions is presented. 
A system of partial differential equations describes the diffusion 
and the binding/unbinding processes in both layers. Additional 
flux continuity at the interface and
clearance conditions into systemic circulation are imposed. An eigenvalue
 problem with discontinuous coefficients is solved and an analytical solution is given
in the form of an infinite series expansion. The model points out the role
of the diffusion and reaction parameters, which control the complex
transfer mechanism and the drug kinetics across the two layers.
Drug masses are given and their dependence on the physical parameters is discussed.
\end{abstract}



\bigskip\bigskip
\section{Introduction}

Transdermal drug delivery (TDD for short) is an approach used to 
deliver drugs through the
 skin for therapeutic purposes as an alternative to oral, intravascular,
 subcutaneous and transmucosal routes. Various TDD 
technologies are possible including the use of suitable drug formulations, 
carriers such as nanoparticles and penetration enhancers to facilitate 
drug delivery and
transcutaneous absorption\footnote{The term ``drug delivery'' refers 
to the release of drug from 
a polymeric platform where it is initially contained. 
The name ``percutaneous absorption'' is generally related to 
the same process viewed from the perspective of the living
tissue where the drug is directed to.  }.
TDD offers several advantages compared to other traditional
delivery methods: controlled release rate, noninvasive administration, less
frequent dosing, and simple application without professional 
medical aids, improving
 patient compliance. For these reasons it represents a valuable and 
attractive alternative to oral administration \cite{chi}.  \par

Drugs can be delivered across the skin to have an effect on the tissues adjacent to the site of application ({\em topical delivery}) or to be effective after distribution through the circulatory system ({\em systemic delivery}). While there are many advantages to TDD, the skin's barrier properties  provide a significant challenge.
To this aim, it is important to understand the mechanism of drug permeation from the delivery  device (or vehicle, typically a 
transdermal patch or medicated plaster, fig. 1) across the 
skin \cite{mit}.  In TDD, the drug
should be absorbed to an adequate extent and rate in order to achieve and maintain
uniform, systemic, effective levels throughout the duration of use.
TDD must be carefully tailored to achieve the optimal therapeutic effect 
 and to deliver the correct
 dose in the required time \cite{pra}. The pharmacological effects of the drug,
tissue accumulation, duration and distribution could potentially have an effect on 
its efficacy and a delicate balance between an adequate amount of drug delivered over an extended period of time 
and the minimal local toxicity should be found \cite{ana}.
Most drugs do not penetrate skin at rates
sufficient for therapeutic efficacy and this restrictive nature limits the use of the transdermal route to molecules of low  molecular weight and with moderate lipophilicity.  In 
general, the first skin layer, the stratum corneum, presents most of the resistance to diffusive
transport into skin. Thus, once the drug molecules cross it, transfer into deeper
dermal layers and systemic uptake occurs in a relatively short time. In order to speed up
transdermal permeation of drugs in the stratum corneum, new delivery techniques are currently under
investigation, for example the use of chemical enhancers or microneedles and techniques 
such as ultrasound, electroporation and iontophoresis \cite{pra, mitra}. \par
Mathematical modelling for TDD
constitutes a powerful predictive tool for the fundamental 
understanding of biotransport 
processes, and for screening processes and stability assessment of new
formulations.  In the absence of experiments, a number of mathematical models
and numerical simulations have been carried out regarding TDD, its efficacy and
 the optimal design of devices  \cite{man1,add,fra, fra1}.
Recent extensive reviews deal with various aspects of 
transdermal delivery at different scales \cite{mit,ana1,pon2, nag}. 
In general, drug absorption into the skin occurs by passive
diffusion and most of the proposed models  
consider this effect only. 
On the other hand, there is a limited effort to explain 
the drug delivery mechanism 
from the vehicle platform.
This is a very important issue indeed, since the polymer matrix acts as a
drug reservoir, and an optimal design of its microstructural characteristics 
would improve the release
performances \cite{rim}.  For example, in the vehicle, the dissolution of the drug from encapsulated to free phase occurs at a given reaction rate. Another relevant feature in TDD
is the similar binding/unbinding process through the receptor sites in the skin.  These drug 
association-dissociaton aspects are often neglected 
or underestimated by most authors who consider purely diffusive systems 
in the skin or in the vehicle \cite{kub1,sim1}. One exception is 
the work of Anissimov et al.\cite{mit,ana}, where a linear reversible binding is considered, but the vehicle is taken 
into account
only through a boundary condition of the first kind. 
However, it is worth emphasizing that 
the drug elution depends on the 
properties  of  the ``vehicle-skin'' system, taken as a whole, and modelled as
a coupled two-layered system.  \par
The method used in the present study follows the mathematical approach 
developed in a series of previously published papers which successfully
describe drug dynamics form an eluting stent embedded in an 
arterial wall [14--17].
In these papers, we proposed a number of models of increasing complexity
to explain the diffusion-advection-reaction release mechanism
of a drug from the stent coating to the wall, constituted of a number of 
contiguous homogeneous media of different properties and extents. 
Separation of variables leads to an eigenvalue problem with
discontinuous coefficients and an exact solution is
given in terms of infinite series expansion  and
 is based on a two- or multi-layer diffusion model. 
In the wake of these papers,  a  two-layer two-phase coupled  model for TDD has been recently
 presented 
and a semi-analytical solution has been proposed for drug concentration and mass in the vehicle and the skin at various times, for special values of 
the parameters \cite{pon1}.
 \par
In the present paper we extend the above study 
and remove some of the simplifying assumptions, obtaining a solution
in a more general form. 
Together with diffusive effects, the drug
dissolution process in the polymer constituting the vehicle 
platform and the reversible
drug binding process in the skin are also addressed.
A solution of the Fick-type reaction-diffusion 
equations (reduced problem) serves 
as the building block to construct a space-time dependent 
solution for the general equations (full
problem).
A major issue in modelling TDD is the assessment of the
key parameters defining skin permeability, diffusion coefficients, drug
dissociation and association rates.
Lacking  experimental data and reliable estimates of the model parameters, we 
carry out a  systematic sensitivity analysis over a feasible range of parameter
values.
The results of the simulations provide valuable insights 
into local TDD and can be used to assess experimental procedures 
to evaluate drug efficacy, for an optimal control
strategy in the
design of technologically advanced transdermal patches.

\section{Formulation of the problem}
\setcounter{equation}{0}
 
To model TDD, let us consider a two-layered system composed 
of: (i) the {\em vehicle} (the transdermal patch or the film of an ointment), and (ii)
 the {\em skin} (the stratum corneum followed by the skin-receptor cells and 
the capillary bed) (fig. 2). 
The drug is stored in the vehicle, a reservoir consisting of a polymeric matrix.
This is enclosed on one side with an impermeable
backing and having on the other side an adhesive in contact with the skin.  
A rate-controlling membrane protecting the polymer matrix may exist.  
In this configuration, the first layer is
shaped as a planar slab that is in direct contact with the skin,
 the second layer. 
As  most of the 
mass dynamics occurs along the direction normal to the 
skin surface, we restrict our study to 
a simplified one-dimensional model. In particular, we consider
as $x$-axis the normal to the skin surface 
 and  oriented with the positive direction outwards the skin.  
Without loss of generality, let $x_0=0$ 
be the vehicle-skin interface and 
$l_0$  and $l_1$ the thicknesses of the vehicle and skin layers respectively
(fig. 2). The vehicle and 
the  skin are both treated from a macroscopic perspective so that they
are represented as two homogeneous media.  \par

Initially, the drug is encapsulated at maximum concentration within the vehicle in a bound phase 
(e.g. nanoparticles or crystalline form) ($c_e$): in a  such state, it is unable to be delivered to the tissue. Then, a fraction of this 
drug ($\beta_0 c_e$) is 
transferred, through an unbinding process, to an unbound -- free, 
biologically available -- phase ($c_0$), and conversely, a part of the free drug ($\delta_0 c_0$) is 
transferred by a binding process to the bound state, in a dynamic equilibrium
(fig. 3). Also, at the same time, another fraction of free drug ($c_1$) begins to
 diffuse into the adjacent skin  ({\em delivery}). 
Similarly,
in the skin -- the release medium --  a part of the unbound drug  ($\beta_1 c_1$) is metabolized by the cell receptors and transformed in a bound state ($c_b$) ({\em absorption}), and with the reverse unbinding
process ($\delta_1 c_b$) again in a unbound phase. Thus, the drug delivery-absorption
 process starts from the vehicle and ends to the skin receptors, with bidirectional phase changes in a cascade sequence, 
as schematically represented
 in fig. 3. Local mass non-equilibrium processes, such as bidirectional drug binding/unbinding phenomena, play a key role in TDD, with characteristic times faster than those of diffusion. In other cases of drug delivery, such as in eluting stents, a second-order saturable reversible binding model has been proposed \cite{tza}:  
this comprehensive model includes a number of drug dependent parameters 
which are difficult to measure experimentally and, nevertheless, does not 
necessarily apply to TDD.  Here, a linear relationship is commonly 
used, as the density of binding sites far exceeds the local free drug concentration
\cite{mit,ana}.  In the first layer the process is
 described by the following equations: 
\begin{align}
& {\p c_e \over \p t} = - 
\beta_0  c_e +   \delta_0 c_0    &\mbox{in} \,  (-l_0,0)\,\,  \label{prob28} \\
& {\p c_0 \over \p t} =D_0 {\p^2 c_0 \over \p x^2} + \beta_0 c_e - \delta_0 c_0      & \mbox{in} \,  (-l_0,0)\,\,   \label{prob29} 
\end{align}
where $D_0$ ($cm^2/s$) is the effective diffusion coefficient of the unbound solute, $\beta_0 \geq 0$ and $\delta_0 \geq 0$ ($s^{-1}$) are the unbinding and binding rate constants in the vehicle, respectively. In detail, the rate of release of encapsulated drug into its free state is implied by the dissociation rate constant
$\beta_0$, while $\delta_0$  provides a representation of the rate at which the
free solute re-associates in the bound state.  \par

Similarly, in the second layer, the drug dynamics 
is governed by similar reaction-diffusion equations:
\begin{align}
& {\p c_1 \over \p t} =D_1 {\p^2 c_1 \over \p x^2} -
  \beta_1 c_1  + \delta_1 c_b  &\mbox{in}& \, \, (0 , l_1)   \label{er1} \\
& {\p c_b \over \p t} = 
  \beta_1 c_1  -  \delta_1 c_b  &\mbox{in}& \, \, (0 , l_1)   \label{er2} 
\end{align}
where $D_1$ is the effective diffusivity of unbound drug, $\beta_1 \geq 0$ and 
$\delta_1 \geq 0$ are the binding and unbinding rate constants in the skin, respectively, defined similarly as above for the vehicle. They can be evaluated experimentally as described in \cite{ana,fra1}, sometimes through the equilibrium dissociation constant $K= \ds{\delta_1 \over \beta_1}$. The magnitudes of $\delta_1$ and  $\beta_1$ are 
inversely proportional 
to the typical times associated with the binding-unbinding processes.  However, these reaction times are 
not negligible compared with the diffusive characteristic times ({\em slow binding}) \cite{pon2}. \par

To close the previous bi-layered mass transfer system 
of eqns. (\ref{prob28})--(\ref{er2}), 
a flux continuity  condition 
has to be assigned at the vehicle-skin interface: 
\be
 -D_0{ \p c_0 \over \p x} = -D_{1}{ \p c_{1} \over \p x}    \qquad \mbox{at} \, \, x=0  \label{eru34} 
\ee

As far as the concentration continuity is concerned, this is not guaranteed because of a different drug partitioning between vehicle and skin. This is taken into account through 
an appropriate mass transfer coefficient $P_r$ \cite{fra1, pon2}. Additionally, a semi-permeable
rate-controlling membrane or an
adhesive film or a non-perfect vehicle-skin contact, having $1 / P_m$ as 
mass resistance, might be present at the interface. Thus, a jump concentration may occur:
\be
-D_1 {\p c_1 \over \p x}= P (c_0-c_1)  \qquad \mbox{at} \, \, x=0   \label{eru35}  
\ee
with $P (cm/s)$ the overall mass transfer coefficient :
\bdm
{1 \over P} = {1 \over P_r} + {1 \over P_m}
\edm
Estimation of the partition coefficient or of its derived quantity 
$P_r$ is a very difficult task. The recent
review of Mitragotri et al. \cite{mit} provides an excellent overview of the current methods used for its representation.
The usually met condition $c_0 \propto c_1$ does not apply, in our opinion, to time dependent cases. \par

No mass flux passes between the vehicle and the external surrounding due 
to the impermeable backing
and we impose a no-flux condition :
\be
D_0{\p c_0 \over \p x} =0  \qquad\qquad \mbox{at $x=-l_0$}  \label{ad1}
\ee

Finally, a boundary condition has to be imposed at the skin-receptor (capillary) boundary. At this point the elimination of drug by capillary system follows first-order kinetics:

\be
  K_{cl} c_1 + D_1 {\p c_1 \over \p x}=0   \qquad\qquad\qquad  \mbox{at    } x=l_1   \label{eqn45}
\ee  
where $K_{cl}$ is the skin-capillary clearance per unit area ($cm/s$). 
The initial conditions are:
\be
c_e(x,0)=C_e  \qquad \qquad c _0(x,0)=0   \qquad \qquad 
   c_1(x,0) = 0    \qquad \qquad  c_b(x,0) = 0   \label{prob20}  
\ee

\subsection{Dimensionless equations}
All the variables and the parameters are now normalized
to get easily computable dimensionless quantities
as follows:
\begin{eqnarray*}
&& \bar x={x \over l_1}  \qquad\qquad   
 \bar t= { D_{1} \over (l_1)^2} t  \qquad\qquad  \phi= { P l_1 \over D_1} 
 \nonumber \\
&&  \bar l_0=   {l_0 \over  l_{1}} \qquad\qquad 
\gamma={D_0 \over  D_{1}}  \qquad\qquad     \bar c_i= {c_i \over C_{e}}      \nonumber \\
&&  K =  {K_{cl} l_1 \over D_1} \qquad\qquad \bar \beta_i= {\beta_i (l_1)^2 \over D_{1}}   \qquad\qquad  \bar \delta_i= {\delta_i (l_1)^2 \over D_{1}} \qquad i=0,1  \label{gh2}
\end{eqnarray*} 
By omitting the bar for simplicity, 
the mass transfer problem (\ref{prob28})--(\ref{er2})  can be now written
in dimensionless form as:

\begin{align}
& {\p c_e \over \p t} = - 
\beta_0  c_e +   \delta_0 c_0    &\mbox{in} \,  (-l_0,0)\,\,  \label{prob38} \\
& {\p c_0 \over \p t} =\gamma {\p^2 c_0 \over \p x^2} + \beta_0 c_e - \delta_0 c_0      & \mbox{in} \,  (-l_0,0)\,\,   \label{prob39}  \\
& {\p c_1 \over \p t} =  {\p^2 c_1 \over \p x^2} -
  \beta_1 c_1  + \delta_1 c_b  &\mbox{in} \, \, (0 , 1)   \label{er10} \\
& {\p c_b \over \p t} = 
  \beta_1 c_1  -  \delta_1 c_b  &\mbox{in} \, \, (0 , 1)   \label{er20} 
\end{align}

and the interface and boundary conditions  (\ref{eru34})--(\ref{eqn45}) read:
\begin{align}
& {\p c_0 \over \p x}=0   &\mbox{at}& \, \, x=-l_0  \nonumber \\
& \gamma { \p c_0 \over \p x} = { \p c_1 \over \p x}      \qquad\qquad		
- { \p c_1 \over \p x}  = \phi ( c_0  - c_1)  &\mbox{at}&  \:\: x=0  \nonumber \\
&  K  c_1 +  { \p c_1 \over \p x}=0    & \mbox{at}&  \, \, x=1  \label{ern47}
\end{align}

supplemented with the initial condition:
\be
c_e(x,0)=1  \qquad \qquad    c_0 (x,0)=0   \qquad \qquad    c_1 (x,0)=0  \qquad \qquad    c_b (x,0)=0     \label{qe6}
\ee
\bigskip

\section{Solving a reduced problem}
\setcounter{equation}{0}

To solve the above problem, let us consider, preliminarily, the associated system 
of P.D.E.'s obtained from (\ref{prob39})--(\ref{er10}) by
setting $c_e=c_b=0$:
\begin{align}
 &     {\p \tilde c_0 \over \p t}  =  \gamma {\p^2 \tilde c_0 \over \p x^2} - \delta_0 \tilde c_0
       \qquad\qquad   -l_0 < x < 0  \nonumber \\
 &     {\p \tilde c_1 \over \p t}  =   {\p^2 \tilde c_1 \over \p x^2} - \beta_1 \tilde c_1
       \qquad\qquad 0< x < 1     \label{reduced}   
   \end{align}   
with the same boundary conditions (\ref{ern47}) and homogeneous initial
conditions. We look for a non trivial solution
by separation of variables:
\be
\tilde c_0(x,t) = X_0(x) G_0(t)   \qquad \qquad  \tilde c_1(x,t)= X_1(x) G_1 (t)
\ee

After substitution, eqns.(\ref{reduced}) become:
\begin{align}
 &     \left( {dG_0 \over dt} + \delta_0 G_0 \right) X_0 = 
              \gamma  X_0^{\prime\prime} G_0   \qquad  -l_0 < x < 0  \nonumber \\
&     \left( {dG_1 \over dt} + \beta_1 G_1 \right) X_1  =
               X_1^{\prime\prime} G_1   \qquad  0 < x < 1   \label{rfg}        \end{align}

From the previous,  $X_0$ and $X_1$ must satisfy the spatial eigenvalue problem:
\begin{align}
 &   X_0^{\prime\prime} = -\lambda_0^2 X_0  \qquad -l_0 < x < 0  \nonumber  \\        
 &   X_1^{\prime\prime} = -\lambda_1^2 X_1  \qquad   0 < x < 1 \label{ty6}
\end{align}

with:
\begin{align}
& X_0'=0   &\mbox{at}& \, \, x=-l_0  \nonumber \\
& \gamma{X_0'} =  {X_1'}      \qquad\qquad		
- { X_1'}  = \phi (X_0  - X_1)  &\mbox{at}&  \:\: x=0  \nonumber \\
&  K  X_1 +  {X'_1}=0    & \mbox{at}&  \, \, x=1  \label{e49}
\end{align}

The eigenfunctions of the problem (\ref{ty6}) are searched as:
\be
X_0(x) = a_0 \cos(\lambda_0 x) + b_0 \sin(\lambda_0 x)
\qquad\qquad
X_1(x)= a_1 \cos(\lambda_1  x) + b_1 \sin(\lambda_1 x) \label{fgh}
\ee
By enforcing the conditions (\ref{e49}), we get the following linear system 
of equations:
\bdm
   \left\{
   \begin{array}{l}
         a_0 \sin(\lambda_0 l_0)  + b_0 \cos(\lambda_0 l_0) = 0
         \\[0.5cm]
         \gamma \lambda_0 b_0 =  \lambda_1 b_1
         \\[0.5cm]
          \phi (a_0 - a_1) +  \lambda_1 b_1 =0
         \\[0.5cm]
         \left[ K \cos(\lambda_1) - \lambda_1 \sin(\lambda_1)\right ] a_1+ 
      \left[ K \sin(\lambda_1) + \lambda_1 \cos(\lambda_1)\right ] b_1=0
   \end{array}
   \right.
\edm

A non trivial solution ($a_0, b_0, a_1, b_1$) with:
\begin{align}
  & a_0= a_1 -{ \lambda_1 \over \phi} b_1  = -b_0 \cot(\lambda_0 l_0)   \nonumber \\
& b_0={\lambda_1 \over \gamma \lambda_0} \: b_1 \nonumber \\
& a_1 = {K \tan \lambda_1 +  \lambda_1  \over 
\lambda_1  \tan \lambda_1 - K } b_1 \label{sol2}
\end{align}
and $b_1$ arbitrary, exists only if the determinant of the coefficient
matrix is zero, i.e.:
\begin{align}
&  \lambda_1 \left(\lambda_1 \tan(\lambda_1) -K  \right)
 \left [\gamma \lambda_0 \tan(\lambda_0 l_0) - \phi  \right]
-\gamma \phi \lambda_0  \tan( \lambda_0 l_0) \left( K \tan(\lambda_1) + \lambda_1   \right)  = 0
\label{det2}
\end{align}
In general, the transcendental eqn (\ref{det2}) admits infinite eigenvalues. On the other hand, from (\ref{rfg}), we have:
\begin{align}
 & {dG_0 \over dt} + (\delta_0 + \gamma \lambda_0^2) G_0 = 0
                \nonumber \\
&  {dG_1 \over dt} + (\beta_1 + \lambda_1^2)  G_1  = 0
                    \label{rfg1}        
\end{align}
In order to satisfy the matching conditions at the interface $x=0$ for all
$t>0$, from (\ref{rfg1}) it follows:
\bdm
G(t) \equiv G_0(t) = G_1(t) = \exp(-\omega t)
\edm
with
\be
\omega \equiv \delta_0 + \gamma \lambda_0^2 = \beta_1  + \lambda_1^2  \label{match}
\ee
From the latter, it follows:
\be
\lambda_0=\sqrt{\lambda_1^2 + \beta_1 -\delta_0 \over \gamma}
\label{edf}
\ee

\noindent and replacing in (\ref{det2}) we get a set 
of eigenvalues ($\lambda_0^k, 
\lambda_1^k$) and of eigenvectors ($X_0^k, X_1^k$) as in eqn. (\ref{fgh}). 
Note
that, although from eqn. (\ref{edf}) some eigenvalues can be imaginary, 
this circumstance is excluded with
the numerical values used (see sect. 5).
We can easily prove that ($X_0^k, X_1^k$) form a orthogonal system,
that is:
\be
\INT^0 _{-l_0}X_0^k X_0^q dx + 
\INT^1_{0} X_1^k X_1^q dx = 
\left\{
\begin{array}{lll}
 0  & \mbox{for} & k \ne q 
\\*[0.25cm]
N_k  & \mbox{for} & k = q
\end{array}
\right. \label{abc}
\ee
where
\begin{eqnarray}
N_k & = & {1 \over 2}   \left[
        \left((a_0^k)^2 + (b_0^k)^2\right) l_0 - {a_0^k b_0^k \over \lambda_0^k}
       +(a_1^k)^2 +1 + {a_1^k  \over \lambda_1^k} \right]    \label{abc1} 
\end{eqnarray}

Finally the concentrations are expressed by summing up all the contributions:
\be
\tilde c_0(x,t) = \SUM_{k=1}^\infty X_0^k(x) G^k (t) \qquad\qquad 
\tilde c_1(x,t) = \SUM_{k=1}^\infty X_1^k(x) G^k (t)  \label{tyu}
\ee

\section{Solution of the full problem}
\setcounter{equation}{0}
 
Consider first that, by making explicit $c_e$ from eqn. (\ref{prob38}), $c_b$ from  eqn. (\ref{er20}),
and from initial conditions (\ref{qe6}), we have
\begin{align}
&c_e(x,t) = \exp(-\beta_0 t) + \delta_0 \INT^t_0 c_0(x,\tau) 
\exp[\beta_0(\tau-t)] d \tau  \label{jkc}
\\*[0.5cm]
&c_b(x,t) =  \beta_1 \INT^t_0 c_1(x,\tau) \exp[ \delta_1(\tau-t)] d \tau
 \label{jkl}
\end{align}
The eqns. (\ref{jkc})--(\ref{jkl}) mean that $c_e$ (resp. $c_b$) are computed in terms 
of  $c_0$ (resp. $c_1$)  and the latter are expressed in the space 
spanned by $X_0^k$ (resp. $X_1^k$) (see eqn. (\ref{fgh})), with the set of eigenvalues and eqn. 
(\ref{match}) unchanged, similarly to eqn. (\ref{tyu}), as:
\begin{align}
c_0(x,t) = \SUM_{k=1}^\infty X_0^k(x) U^k (t) \qquad\qquad
c_1(x,t) = \SUM_{k=1}^\infty X_1^k(x) U^k (t)  \label{mn1} 
\end{align}
where the time functions $U^k(t)$ have to be computed for the complete system (\ref{prob38})--(\ref{er20}).
Replacing (\ref{mn1}), eqns. (\ref{jkc})-(\ref{jkl}) are written as:
\be
c_e(x,t) =  \exp(-\beta_0 t) + \delta_0 \SUM_{k=1}^\infty 
X_0^k(x) H_0^k(t) \qquad\qquad
c_b(x,t) =  \beta_1 \SUM_{k=1}^\infty X_1^k(x) H_1^k(t)
\label{decog}
\ee
with
\be
H_0^k(t) =  \INT_0^t U^k(\tau) \exp[\beta_0(\tau-t)] d \tau \qquad\qquad
H_1^k(t) = \INT_0^t U^k(\tau) \exp[\delta_1(\tau-t)] d \tau \label{kl1}
\ee

\noindent By inserting (\ref{mn1})--(\ref{decog}) into eqn. (\ref{prob39}), 
multiplying  by $X_0^p$, we get:
\begin{align}
&\SUM_{k=1}^\infty X_0^k X_0^p {dU^k \over dt}= -\gamma 
\SUM_{k=1}^\infty X_0^k X_0^p (\lambda_0^k)^2 U^k  
-\delta_0 \SUM_{k=1}^\infty X_0^k X_0^p U^k  \nonumber \\
& +\beta_0 \exp (-\beta_0 t) X_0^p + \beta_0 \delta_0 
\SUM_{k=1}^\infty X_0^k X_0^p H_0^k
\end{align}
Similarly, multiplying eqn. (\ref{er10})  by $X_1^p$, we have:
\be
\SUM_{k=1}^\infty X_1^k X_1^p {dU^k \over dt}= -
\SUM_{k=1}^\infty X_1^k X_1^p (\lambda_1^k)^2 U^k  
-\beta_1 \SUM_{k=1}^\infty X_1^k X_1^p U^k  
 + \beta_1 \delta_1 \SUM_{k=1}^\infty X_1^k X_1^p H_1^k
\ee

Integrating  the previous eqns over the corresponding layers 
and summing up, we get:

\begin{align}
&\SUM_{k=1}^\infty {dU^k \over dt} \left( \INT_{-l_0}^0 X_0^k X_0^p dx 
+  \INT_{0}^1 X_1^k X_1^p dx \right) = -
\SUM_{k=1}^\infty  U^k \left( \gamma (\lambda_0^k)^2 \INT_{-l_0}^0 X_0^k X_0^p dx 
+ (\lambda_1^k)^2  \INT_{0}^1 X_1^k X_1^p dx \right) \nonumber \\
&-\delta_0 \SUM_{k=1}^\infty   U^k \INT_{-l_0}^0 X_0^k X_0^p dx 
-\beta_1 \SUM_{k=1}^\infty   U^k \INT_{0}^1 X_1^k X_1^p dx   \nonumber \\
&+ \beta_0 \exp(-\beta_0 t)  \INT_{-l_0}^0 X_0^p dx  + \beta_0 \delta_0 \SUM_{k=1}^\infty   H_0^k \INT_{-l_0}^0 X_0^k X_0^p dx  + \beta_1 \delta_1 \SUM_{k=1}^\infty   H_1^k \INT_{0}^1 X_1^k X_1^p dx  \label{bn1}
\end{align}

We now pose:
\be
\theta^p_0 = \INT_{-l_0}^0 X_0^p(x) dx= - {b_0^p \over \lambda_0^p} 
\ee
\be
\alpha_0^{kp} = \INT_{-l_0}^0 X_0^p(x) X_0^k(x)  dx \qquad\qquad
\alpha_1^{kp} = \INT_0^1 X_1^p(x) X_1^k(x)  dx
\ee

Note that the space integrated constants $\theta^p_0$ and $\alpha_0^{kp}$ are the same as those computed in Ref. \cite{pon3}, and, from the orthogonality condition (\ref{abc}), they satisfy: 
\be
 \alpha_0^{kp} + \alpha_1^{kp} = 
\left\{
\begin{array}{lll}
 0  & \mbox{for} & k \ne p 
\\*[0.25cm]
N_k  & \mbox{for} & k = p
\end{array}
\right.
\ee
By means of eqn. (\ref{match}), the manipulation of 
eqn. (\ref{bn1}) yields:
\be
\begin{array}{l}
\ds{dU^p \over dt} + \omega^p U^p = \frac{1}{N_p} \left(
\beta_0 \exp(-\beta_0 t) \theta^p_0 
+ \beta_0 \delta_0 \SUM_{k=1}^\infty H_0^k \alpha_0^{kp} 
+ \beta_1 \delta_1  \SUM_{k=1}^\infty H_1^k \alpha_1^{kp}  
\right)
\end{array}
\label{evog}
\ee

From eqn (\ref{kl1}), $H_0^p(t)$ and $H_1^p(t)$ can be computed via the Leibnitz rule as: 
\be
\begin{array}{l}
\ds{dH_0^p \over dt} + \beta_0 H_0^p =  U^p 
\\*[0.5cm]
\ds{dH_1^p \over dt} + \delta_1 H_1^p = U^p 
\end{array}
\label{evoh1}
\ee

The system of the three ODE's 
 (\ref{evog})--(\ref{evoh1}), with homogeneous initial conditions:
\be
U^p(0) = H_0^p(0) = H_1^p(0) = 0  \label{lop}
\ee
is solved numerically  with an explicit Runge-Kutta type solver with an adaptive time step. The
obtained functions $U^p$, $H_0^p$ and $H_1^p$ allow the computation
of all concentrations in eqns. (\ref{mn1})--(\ref{decog}).


\bigskip
From the analytical form of the solution given by 
eqns. (\ref{decog})  the  drug masses are easily computed as integrals 
of the concentrations over the correspondent layer:
\begin{align}
 &M_j(t)= \int\limits  c_j(x,t) dx  \qquad \qquad j=e,0,1,b
\end{align}

Furthermore, the fraction of drug mass retained in each layer and phase
 is computed as:
\be
\mu_j(t)= {M_j(t) \over M_e(0)}   \qquad \qquad j=e,0,1,b  \label{vg1}
\ee 
These are useful indicators of the drug released, diffused and absorbed 
during time.

\section{Numerical simulations and results}
\setcounter{equation}{0}
A common difficulty in simulating physiological processes 
is the identification of reliable estimates of the model parameters. 
Experiments of TDD are impossible or
prohibitively expensive in vivo and the only available source
are lacking and incomplete data from literature. 
The physical problem depends on a large number of parameters, each of them may vary in a finite range, with  a variety of combinations and limiting cases.
The model constants cannot be chosen independently from each other and there is a compatibility condition among them.  
In this paper, for simplicity, the following physical parameters are kept fixed for simulations in TDD
\cite{ana,pon2,kub1,sim1}:
\be
  D_0=  5 \cdot 10^{-7} cm^2 /s   \qquad  D_1= 7 \cdot 10^{-8} cm^2/s  \qquad  P= 10^{-6} cm/s  
\qquad K_{cl} = 3 \cdot 10^{-3} cm/s  \label{eqr}
\ee
and the binding/unbinding parameters are varied to study the effect and to
quantify the sensitivity of the delivery system, with the condition that 
the characteristic reaction times are smaller than the diffusion times, i.e.:
$\beta_0 > \ds{D_0 \over l_0^2}$ and $\beta_1  > \ds{D_1 \over l_1^2}$.

The thickness of the vehicle is set as $l_0 =40 \mu m$, whereas the 
limit of the skin layer ($l_1$) is estimated by 
the following considerations. 
Strictly speaking, in a diffusion-reaction problem the concentration vanishes
 asymptotically at infinite distance. However, for computational purposes, 
the concentration is damped out (within a given tolerance) 
over a finite distance at a given time. Such a length (named {\em penetration distance}, see \cite{pon}) critically depends on the diffusive properties of the two-layered medium and, in particular, is related to the ratio $\ds{D_0 \over D_1}$. In our case,  $l_1$ falls beyond the stratum corneum thickness, say  $l_1 = 0.1 cm$.
 
All the series appearing in the solution (\ref{decog}) and following, have been truncated at a number of 40 terms. \par
\medskip

The concentration profiles are almost flat
in the vehicle, because of its small size, 
are discontinuous at the interface, and
have a space decreasing behavior at any time 
in the skin layer, damping out within the penetration 
distance at all times (fig. 4).   
In the skin, a fast phase exchange of drug occurs at early times, more 
evident
in regions close to the interface $x=0$, and continues at later times (fig. 5).
The mass $M_e$ is decreasing in the vehicle, and $M_0$ (resp. $M_1, M_b$), is first increasing up to 
some upper bound $M_0^*$ (resp. $M_1^*, M_b^*$) (at a time $t_0^*$ (resp. $t_1^*, t_b^*$), with
 $t_0^* < t_1^* < t_b^*$)
and then decaying asymptotically with time (fig. 6).

The effect of binding/unbinding is studied by varying systematically the values of the on-off
 reaction rates  $\beta_0, \beta_1, \delta_0, \delta_1$ over an extended range (this is made
feasible by setting the other parameters as in eqn. (\ref{eqr})).
One parameter is changed at a time, letting the others fixed.  
The occurrence and the magnitude 
of the drug peak as well as the time scale of the absorption 
process result to be very sensitive
 to the mutual size of reaction parameters, combined with those of diffusive
coefficients. In tables 1-4 these values are reported for a number of cases.\\

 \underline{Effect of $\beta_0$ and  $\delta_0$ (tables 1 and 2).} \\
Small values of $\beta_0$ make the dissolution process slower. 
In the limit  $\beta_0 \rightarrow 0$ all drug tends to remain in the phase $c_e$ and it is hardly 
released and absorbed. For $\beta_0$ large, the phenomenon
 is characterized by marked
peaks at early instants. 
On the other hand, for a fixed  $\beta_0$, the TDD can be
 greatly influenced by the possible drug re-association.  
For a large $\delta_0$ the release is slowed down and the drug levels appear  
to be more uniform, with lower peak values.\\

\underline{Effect of  $\beta_1$ and $\delta_1$ (tables 3 and 4).}  \\
In the second layer, the variation of $\beta_1$ and $\delta_1$  influences
only the mass $M_1$ and especially $M_b$, letting the dynamics of $c_e$ and $c_0$ unchanged. 
A small $\beta_1$ is responsible for a raise of $M_1$, leaving  $M_b$ small.
 For a $\beta_1$ sufficiently large (in our simulations 
$\approx 2 \cdot 10^{-4} s^{-1}$) 
bound drug tends to accumulate at a much greater rate than it can be transferred from the contiguous layer and the process cannot be sustained by this value. 
  For small values of  $\delta_1$ the replenishment of the layer is much faster. In the limit $\delta_1 \rightarrow 0$, $M_1$ vanishes after a short transient and $M_b$ reaches a steady value. 
\medskip 

These outcomes provide valuable indicators 
to assess  whether drug reaches a target tissue, 
and to optimize the dose capacity in the vehicle. It appears that 
the relative size of the binding/unbinding parameters
 affects the drug transfer processes, thus influencing the mechanism
 of the whole dynamics. For example, tables 1-4 show which 
set of parameters guarantees a more prolonged 
and uniform release and what other values are 
responsible for a localized peaked distribution followed by a faster decay. 
Thus, the benefit of reaching the desired delivery rate is obtained 
with a proper choice of the physico-chemical-geometrical parameters. 
The resistance of skin to diffusion has to be reduced in order to allow
drug molecules to penetrate and maintain therapeutic levels for an extended period of time. Increasing skin permeability is a prerequisite 
for successful delivery of new macromolecular drugs and improved 
delivery of conventional drugs. 
 The present TDD model constitutes a simple tool that can help in  designing and in manufacturing new vehicle platforms that guarantee the optimal release for an extended period of time.

\section{Conclusions}

Currently TDD is one of the most promising method for drug 
administration and an increasing number of drugs are being added 
to the list of therapeutic agents that can be delivered topically 
or systemically through the skin. 
A deeper understanding of drug
release  is necessary for a rational design of TDD system to
optimize therapeutic efficacy and minimize local toxicity. 
It is important to  find a delicate balance between achieving a highly 
effective result without compromising the safety of the patient. 
One of the approaches to evaluate the characteristics of drug elution from 
the transdermal patch 
 into the skin and to optimize the physico-chemical parameters 
is the mathematical modelling and the numerical simulation. \par
This paper describes the dissolution and the kinetics of a drug 
in the delivery device together with  
the percutaneous absorption in the skin, as a unique system.
This is accomplished by developing a
concentration closed-form solution of a two-phase two-layer model.
The analytical approach is useful  for experimental design and clinical
application, providing the basis for the optimization of parameters. It helps
in identify and quantify, among the others, the relevant concurrent 
effects in TDD. 
The approach
captures the essential physics of drug release 
and dynamics of the percutaneous absorption. The methodology
is equally applicable to other delivery systems, such as
the drug-eluting stent.
\bigskip\bigskip \\
\noindent \underline{Acknowledgments} \\
We are grateful to  A. Di Mascio and S. McGinty for their valuable
discussions and helpful comments. 
This study was partially supported by the MIUR-CNR project ``Interomics'', 2013.

\newpage

\begin{figure}[ht!]
\centering\includegraphics[width=0.5\textwidth]{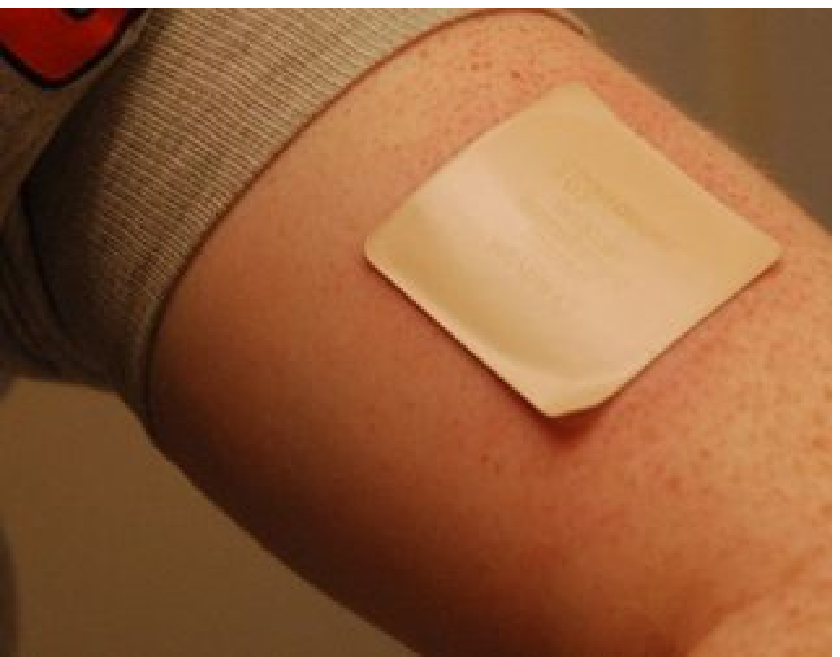} 
\caption{The transdermal patch, a typical vehicle in transdermal drug delivery.}  
\end{figure}
\bigskip\bigskip\bigskip\bigskip

\begin{figure}[ht!]
\centering\scalebox{0.9}{\includegraphics{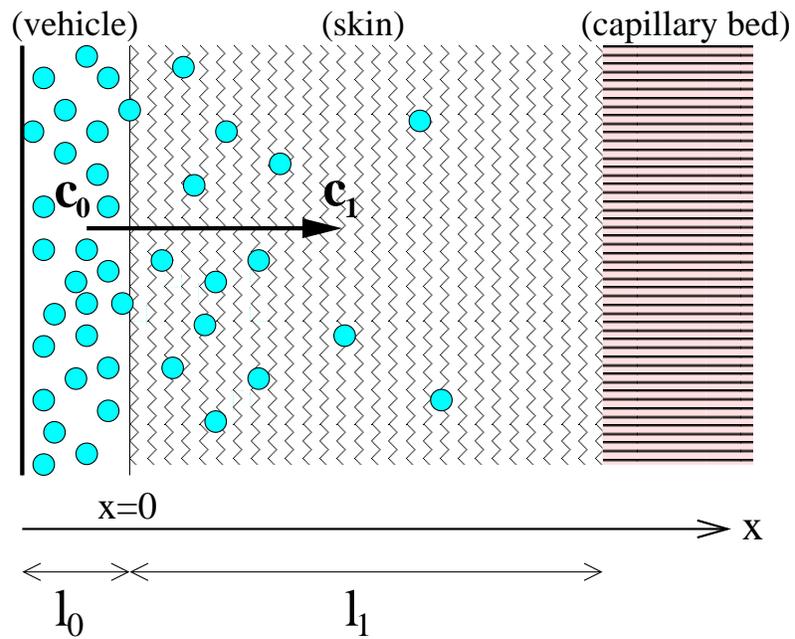}} 
\caption{Cross-section of the vehicle and the skin layers, 
geometrical configuration and reference system in TDD. 
Due to an initial difference of free drug concentrations $c_0$ and $c_1$, a mass flux is established at the interface and
  drug diffuses through the skin.  At a distance
$x=l_1$ the skin-receptor (capillary bed) is present where all drug is 
assumed to be absorbed. Figure not to scale.}  
\end{figure}

\vspace{8mm}
\begin{figure}[ht!]
\centering\scalebox{0.5}{\includegraphics{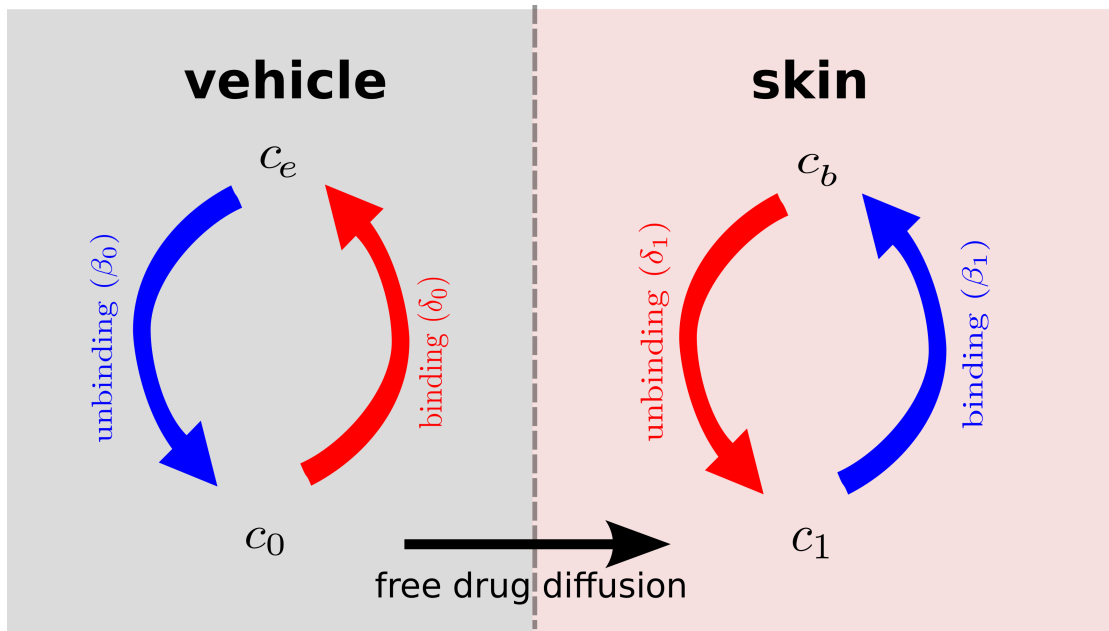}} 
\caption{A diagram sketching the cascade mechanism of drug
delivery and percutaneous absorption in the vehicle-skin coupled system. 
A unbinding (resp. binding) reaction  occurs in the vehicle (resp. in the skin)
(blue arrows). In both layers, reverse reactions (red arrows) 
are present in a dynamic equilibrium. 
Drug diffusion occurs only in the free phases $c_0$ and $c_1$.}  
\end{figure}

 \begin{figure}[ht] 
\bc
\includegraphics[width=0.45 \textwidth]{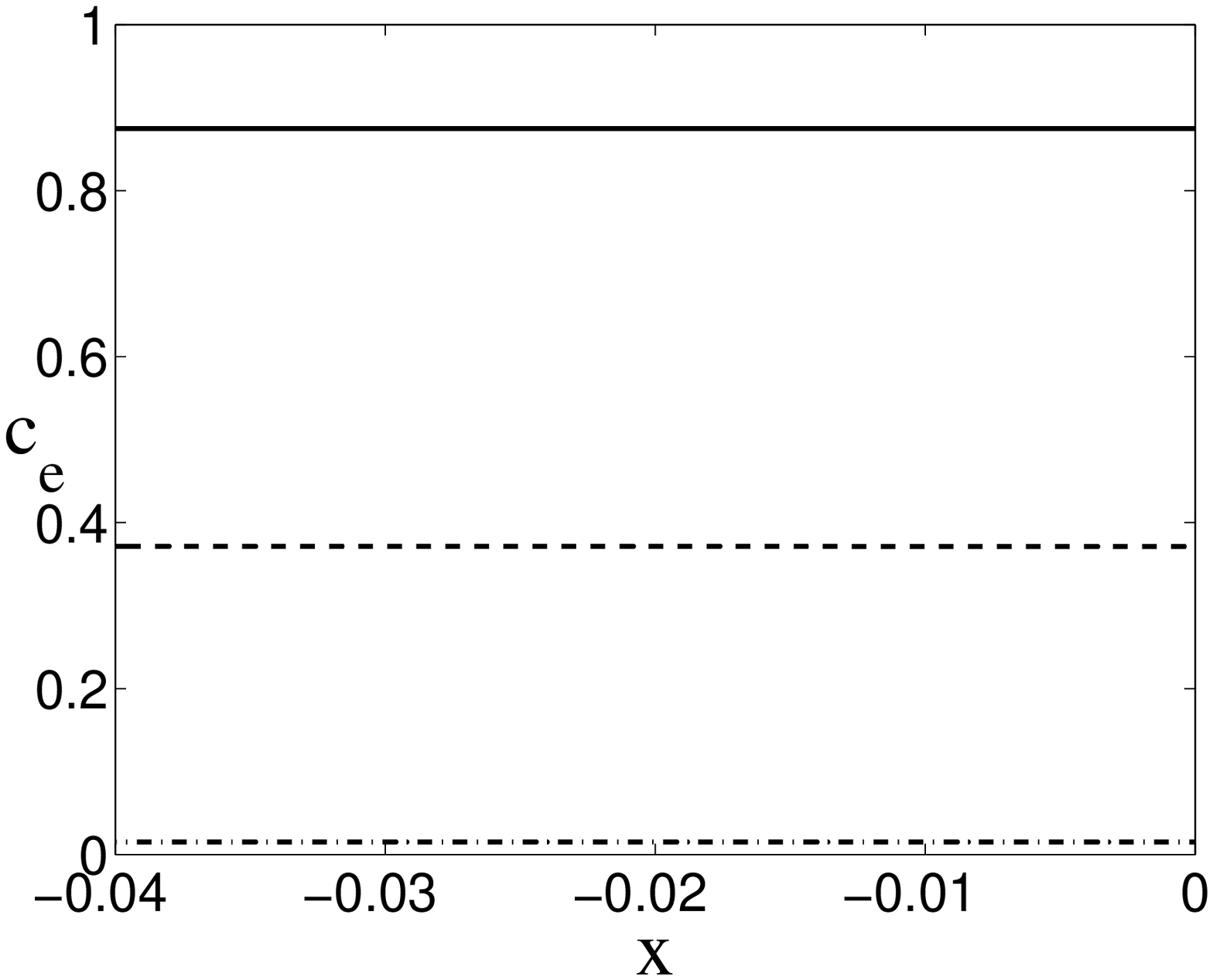}  \qquad
\includegraphics[width=0.45 \textwidth]{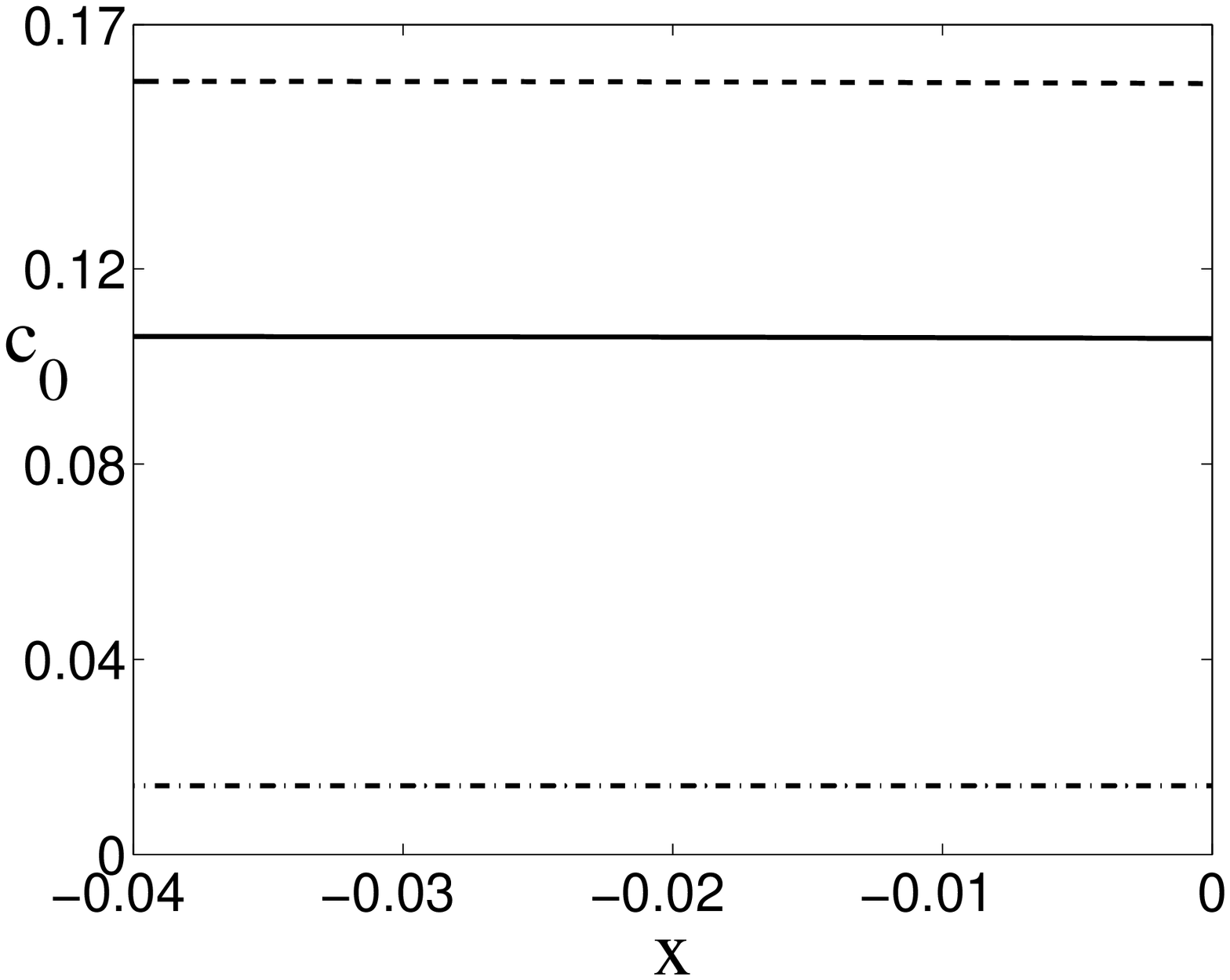} \vspace{4mm} \\
\includegraphics[width=0.45 \textwidth]{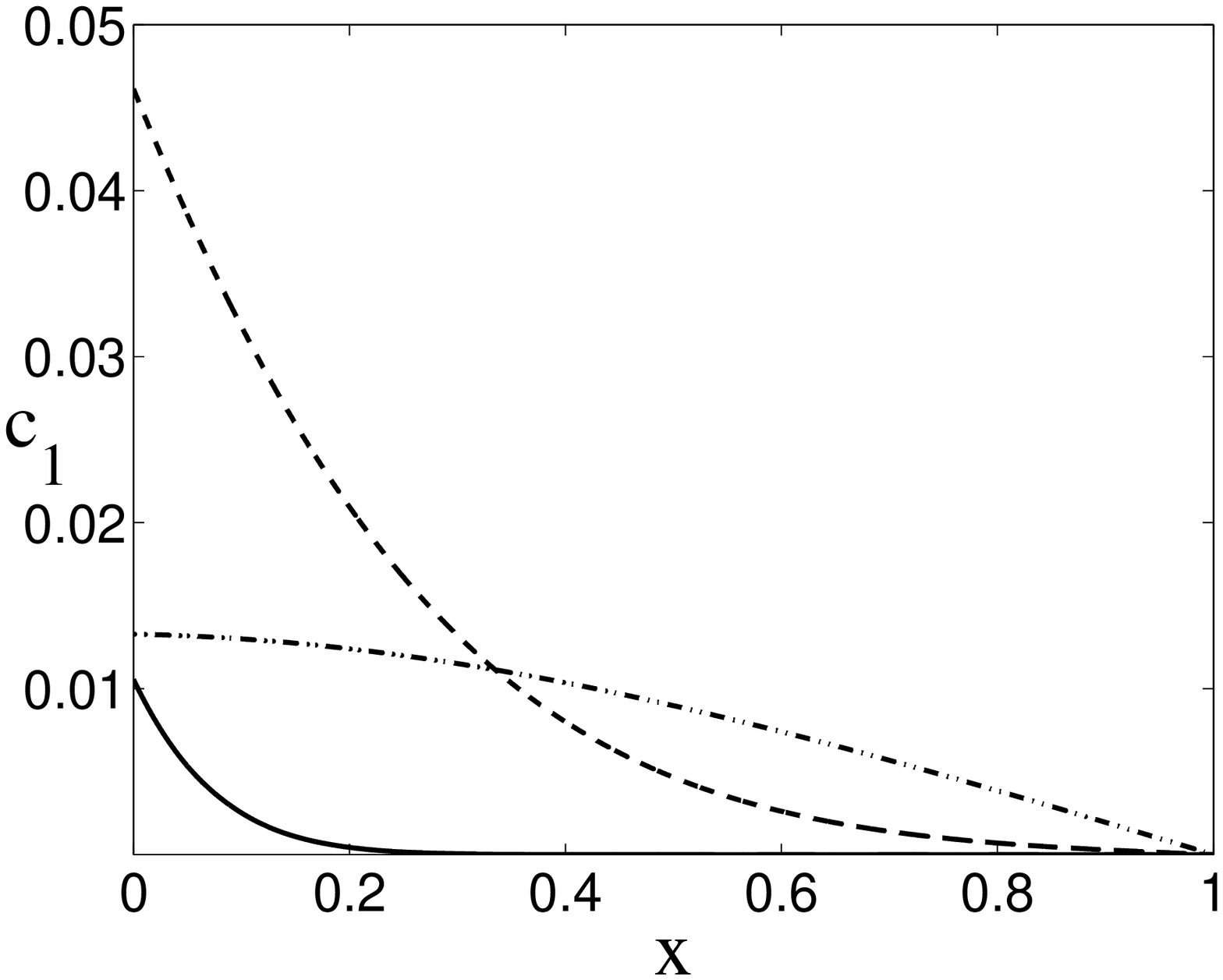} \qquad 
\includegraphics[width=0.45 \textwidth]{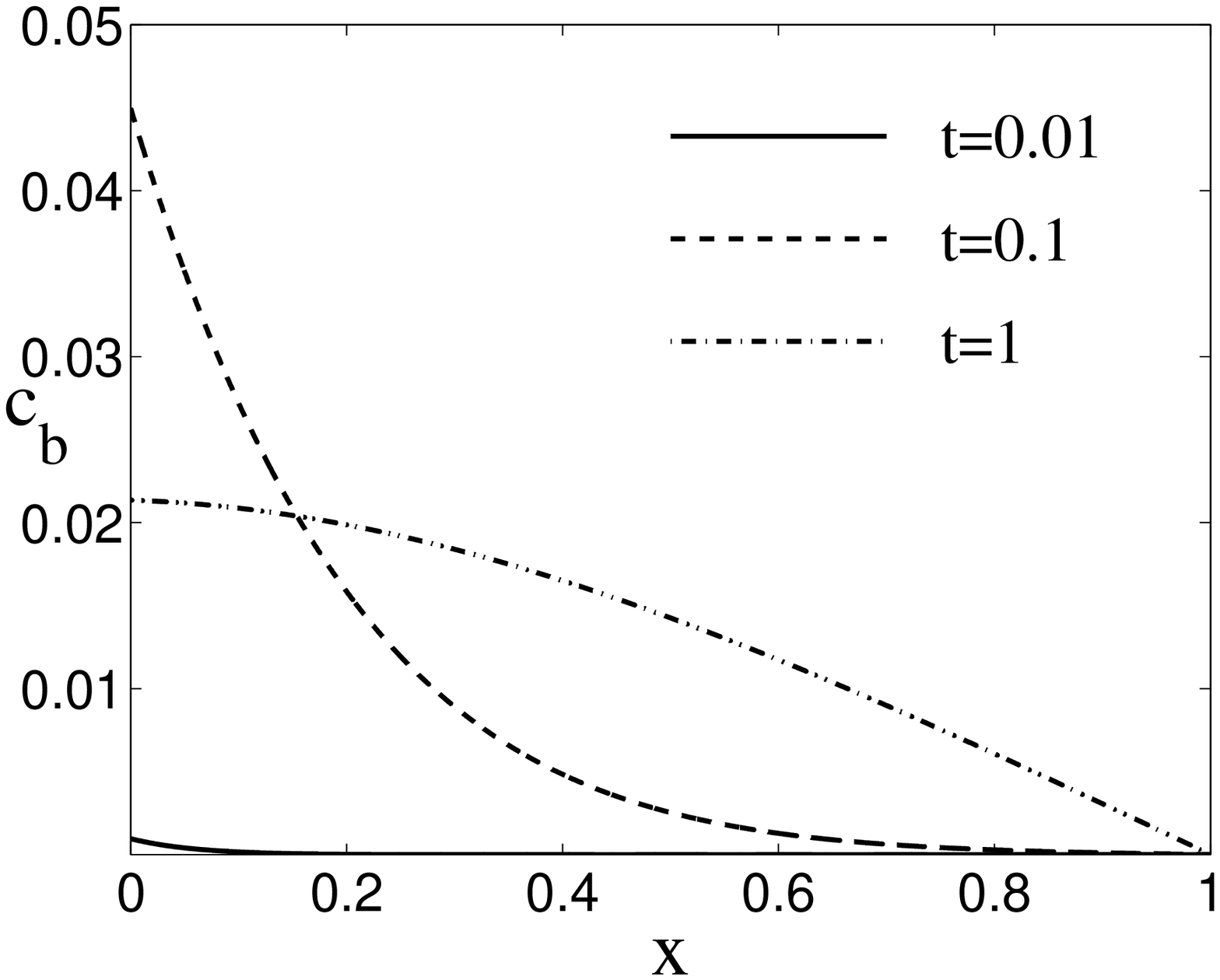}
\ec
\caption{Concentration profiles  in the vehicle ($c_e$ and $c_0$ above) and 
in the skin  ($c_1$ and $c_b$ below)  for the following 
dimensional binding/unbinding 
parameters ($s^{-1}$):
$\beta_0=10^{-4}, \beta_1=1.5 \cdot 10^{-4},
  \delta_0=10^{-4}, \delta_1=10^{-4}$, at three times (note the different scale of concentrations between the two layers).}
\end{figure}

\begin{figure}[ht] 
\centering\scalebox{0.6}{\includegraphics{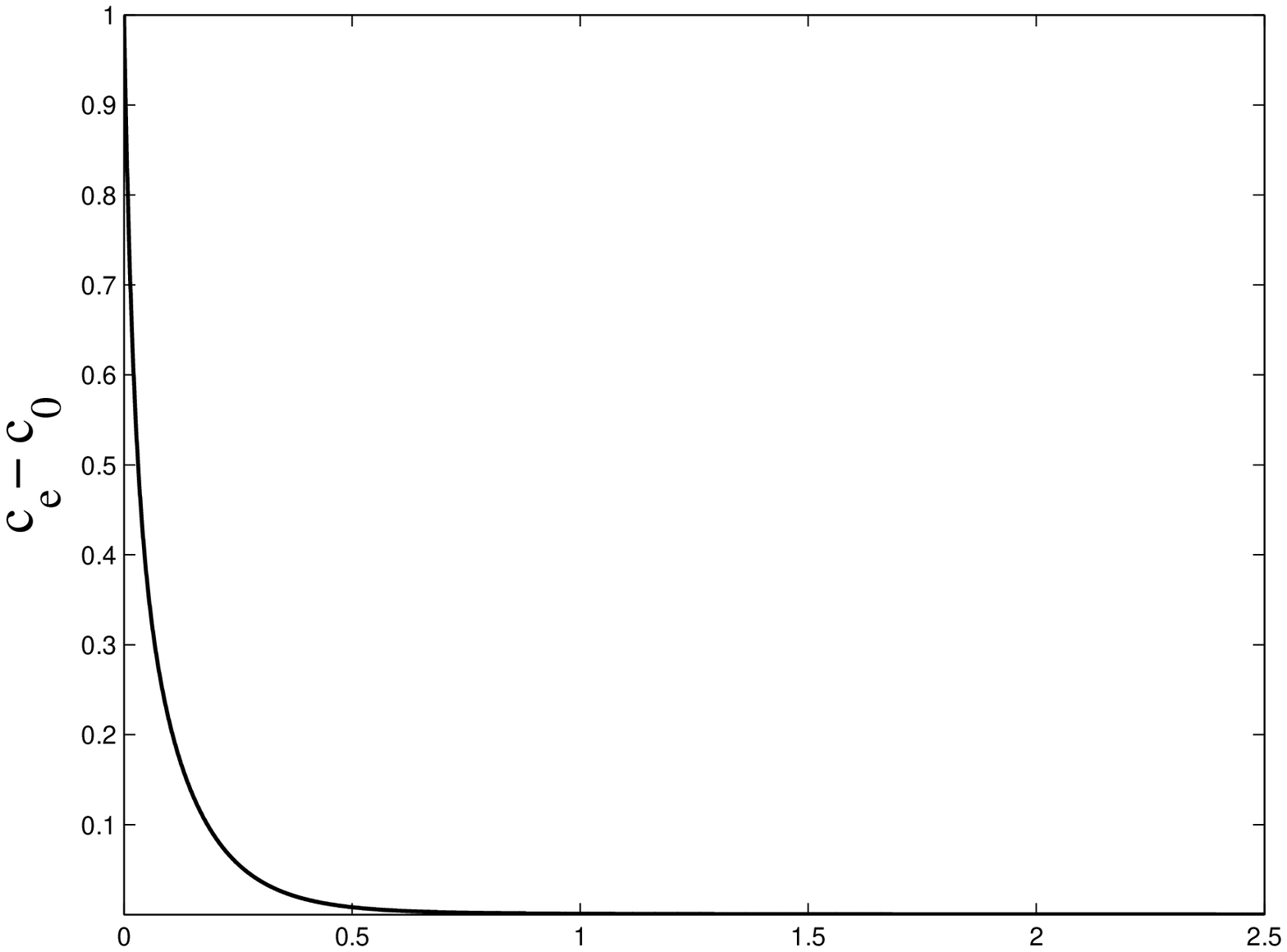}} \\
\hspace{-3mm}
\centering\scalebox{0.6}{\includegraphics{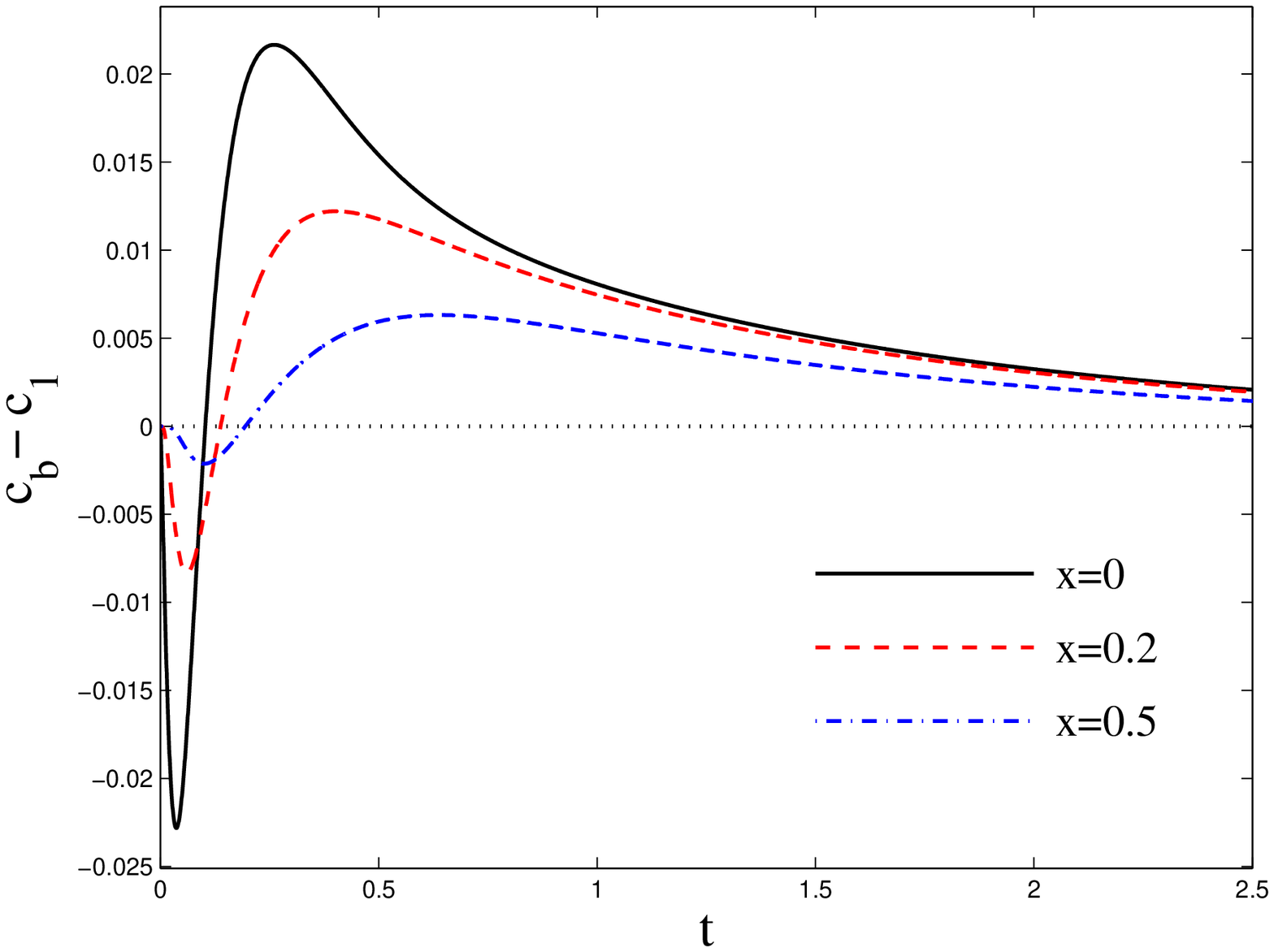}}
\caption{Difference between bound and free concentrations $c_e - c_0$ (top) and  $c_b - c_1$ (bottom) as a function of time
for the same binding-unbinding parameters of fig. 4. In the vehicle a fast decaying phase transfer is evidenced. Due to its thin size, this phase transition $c_e \rightarrow c_0$ occurs at the same manner at any location, whereas in the skin it is shown the larger amount of drug phase change at locations near the interface $x=0$. 
}
\end{figure}

 \begin{figure}[ht] 
\centering\scalebox{0.8}{\includegraphics{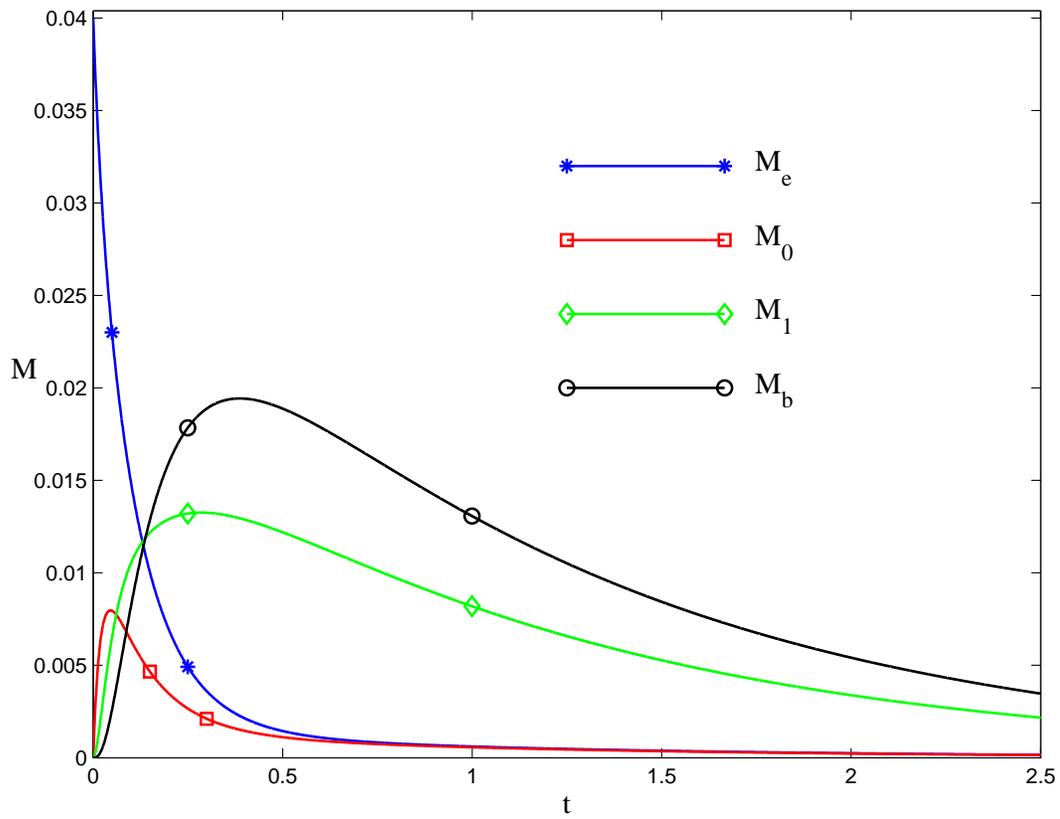}}
\caption{Time histories of the drug mass in the vehicle ($M_e$ and $M_0$)
and in the skin ($M_1$ and $M_b$) for the same binding-unbinding parameters of fig. 4. Only $M_e$ exhibits an exponential decay, whereas $ M_0, M_1$ and $M_b$  increase at initial times, reach a peak $M^*$ at time $t^*$, and then damp to zero at a given rate.}
\end{figure}

\begin{table}[ht]
\caption{Percentage of the two drug mass phases retained in each layer  at different 
times for dimensional bindind/unbinding 
parameters ($s^{-1}$): $\beta_1=10^{-4},  \delta_0=10^{-4}, 
\delta_1=10^{-4} $  and varying 
$\beta_0$ (cfr. eqn. (\ref{vg1})). In the last row of each block the quantity $t^* \h M^*$ refers to the maximum values of 
$M^*$ and the correspondent  times $t^*$ where they are attained.
}
\bc
\begin{tabular}{cllcccc}
\hline
$\beta_0 \,(s^{-1})$ & $t (-)$ & $ t (d:h:m)$ & $\mu_e (\%) $ & $\mu_0 (\%) $ & $\mu_1 (\%)$ & $\mu_b (\%) $ \\
\hline
\multirow{4}*{$10^{-6}$} 
&$0.01$ &  $\simeq 24m$ & $99$ &   $0.1$ & $0.01$  &  $<0.01$    \\ 
&$0.05$  & $\simeq 2h$ & $99$   & $0.2$   &  $0.2$  &  $0.05$  \\ 
&$0.5$  & $\simeq 19h:50m$ & $95$ &  $0.4$ & $1.8$ & $1.6$  \\ 
&$5 $ &  $\simeq 8d:6h:24m$ & $68$ &  $0.3$ & $ 2.6$ & $2.6$  \\ 
$t^* \h M^*$ & & & & $1.32 \h 1.7 \cdot 10^{-4}$ &  $2.06 \h 1.2 \cdot 10^{-3}$  & $2.13 \h 1.2 \cdot 10^{-3}$ \\
\hline
\multirow{4}*{ $ 10^{-4}$ }
&$0.01$ & $\simeq 24m$ & $87$ &   $10$ & $1.8$  &  $0.09$    \\ 
&$0.05$  &  $\simeq 2h$ & $57$   & $19$   &  $18$  &  $4.7$  \\ 
&$0.5$  &   $\simeq 19h:50m$ & $3.9$ &  $3.0$ & $34$ & $36$  \\ 
&$5 $ &  $\simeq 8d:6h:24m$  & $0.01$ &  $0.01$ & $ 0.2$ & $0.2$  \\ 
$t^* \h M^*$ & & & & $0.04 \h 8 \cdot 10^{-3}$ & 0.26$\h$0.015 & 0.36$\h$0.015 \\
\hline
\multirow{4}*{$10^{-2}$} 
&$0.01$ &  $\simeq 24m$ & $0.75$ &   $73$ & $23$  &  $1.6$    \\ 
&$0.05$  &   $\simeq 2h$ & $0.2$   & $25$   &  $53$  &  $20$  \\ 
&$0.5$  &   $\simeq 19h:50m$& $0.02$ &  $2.2$ & $32$ & $35$  \\ 
&$5 $ &  $\simeq 8d:6h:24m$ & $<0.01$ &  $0.01$ & $0.2$ & $0.2$  \\ 
 $t^* \h M^*$   & & & & 0.002$\h$0.036 & 0.06$\h$0.021 & 0.2$\h$0.018 \\
\hline
\end{tabular}
\ec
\end{table}

\begin{table}[ht]
\caption{As in table 1,  for  $\beta_0=10^{-4},  \beta_1=10^{-4}, \delta_1=10^{-4} \, (s^{-1}) $  and varying $\delta_0$ (cfr. with second case of table 1).}
\bc
\begin{tabular}{cllcccc}
\hline
$\delta_0 \,(s^{-1})$ & $t (-)$ & $ t (d:h:m)$ & $\mu_e (\%) $ & $\mu_0 (\%) $ & $\mu_1 (\%)$ & $\mu_b (\%) $ \\
\hline
\multirow{4}*{$10^{-8}$} 
&$0.01$ & $\simeq 24m$ & $86$ &   $11$ & $1.9$  &  $0.09$    \\ 
&$0.05$  & $\simeq 2h$  & $48$   & $24$   &  $21$  &  $5.4$  \\ 
&$0.5$  & $\simeq 19h:50m$ & $0.07$ &  $2.5$ & $35$ & $37$  \\ 
&$5 $ &  4d:17h & $<0.01$ &  $0.01$ & $ 0.2$ & $0.2$  \\ 
$t^* \h M^*$ & & & & $0.05 \h 9.8 \cdot 10^{-3}$ &  $0.19 \h 0.017 $  & $0.31 \h 0.016$ \\
\hline
\multirow{4}*{ $ 10^{-2}$} 
&$0.01$ & $\simeq 24m$& $98$ &   $0.9$ & $0.2$  &  $0.01$    \\ 
&$0.05$  &  $\simeq 2h$& $97$   & $0.9$   &  $1.0$  &  $0.3$  \\ 
&$0.5$  &  $\simeq 19h:50m$ & $88$ &  $0.8$ & $4.5$ & $4.1$  \\ 
&$5 $ &   $\simeq 8d:6h:24m$ & $47$ &  $0.4$ & $ 3.6$ & $3.6$  \\ 
$t^* \h M^*$ & & & & $0.006 \h 3.8 \cdot 10^{-4}$   & $1.30 \h  2.5 \cdot 10^{-3}$  & $1.37 \h 2.5 \cdot 10^{-3}$ \\
\hline
\end{tabular}
\ec
\end{table}

\begin{table}[ht]
\caption{As in table 1, for  $\beta_0=10^{-4},  \delta_0=10^{-4}, \delta_1=10^{-4} \, (s^{-1})$  and varying  $\beta_1$ (cfr. with second case of table 1).}
\bc
\begin{tabular}{cllcccc}
\hline
$\beta_1 \,(s^{-1})$ & $t (-)$ & $ t (d:h:m)$ & $\mu_e (\%) $ & $\mu_0 (\%) $ & $\mu_1 (\%)$ & $\mu_b (\%) $ \\
\hline
\multirow{4}*{$10^{-6}$} 
&$0.01$ &  $\simeq 24m$& $87$ &   $10$ & $1.9$  &  $<0.01$    \\ 
&$0.05$  &  $\simeq 2h$ & $57$   & $20$   &  $22$  &  $0.05$  \\ 
&$0.5$  & $\simeq 19h:50m$ & $4.9$ &  $3.9$ & $48$ & $0.5$  \\ 
&$5 $ &   $\simeq 8d:6h:24m$ & $<0.01$ &  $<0.01$ & $ <0.01$ & $<0.01$  \\ 
$t^* \h M^*$ & & & & $0.04 \h 8 \cdot 10^{-3}$ &  $0.24 \h 0.026$  & $0.33 \h 2.4 \cdot 10^{-4}$ \\
\hline
\multirow{4}*{ $ 2 \cdot 10^{-4}$} 
&$0.01$ & $\simeq 24m$ & $87$ &   $10$ & $1.7$  &  $0.18$    \\ 
&$0.05$  &  $\simeq 2h$ & $57$   & $19$   &  $14$  &  $8.2$  \\ 
&$0.5$  &  $\simeq 19h:50m$ & $3.3$ &  $2.5$ & $26$ & $55$  \\ 
&$5 $ &   $\simeq 8d:6h:24m$ & $0.06$ &  $0.06$ & $ 0.9$ & $2.0$  \\ 
$t^* \h M^*$ & & & & 0.04$\h 8 \cdot 10^{-3}$  & 0.30$\h$0.011 & 0.40$\h$0.022 \\
\hline
\end{tabular}
\ec
\end{table}

\begin{table}[ht]
\caption{As in table 1, for  $\beta_0=10^{-4},  \delta_0=10^{-4}, \beta_1=10^{-4} \, (s^{-1})$  and varying  $\delta_1$ (cfr. with second case of table 1).}
\bc
\begin{tabular}{cllcccc}
\hline
$\delta_1 \,(s^{-1})$ & $t (-)$ & $ t (d:h:m)$ & $\mu_e (\%) $ & $\mu_0 (\%) $ & $\mu_1 (\%)$ & $\mu_b (\%) $ \\
\hline
\multirow{4}*{$10^{-8}$} 
&$0.01$ & $\simeq 24m$  & $87$ &   $10$ & $1.8$  &  $0.09$    \\ 
&$0.05$  &  $\simeq 2h$ & $57$   & $19$   &  $17$  &  $5.6$  \\ 
&$0.5$  & $\simeq 19h:50m$ & $1.7$ &  $0.8$ & $2.4$ & $90$  \\ 
&$5 $ &   $\simeq 8d:6h:24m$ & $<0.01$ &  $<0.01$ & $ <0.01$ & $95$  \\ 
$t^* \h M^*$ & & & & $0.04 \h 8 \cdot 10^{-3}$ &  $0.11 \h 0.01$  & $1.62 \h 0.03$ \\
\hline
\multirow{4}*{ $ 10^{-2}$} 
&$0.01$ & $\simeq 24m$ & $87$ &   $10$ & $1.9$  &  $0.01$    \\ 
&$0.05$  &  $\simeq 2h$ & $57$   & $20$   &  $22$  &  $0.2$  \\ 
&$0.5$  &  $\simeq 19h:50m$ & $4.9$ &  $3.9$ & $48$ & $0.4$  \\ 
&$5 $ &   $\simeq 8d:6h:24m$  & $<0.01$ &  $<0.01$ & $ <0.01$ & $<0.01$  \\ 
$t^* \h M^*$ & & & & $0.04 \h 8 \cdot 10^{-3}$ & 0.24$\h$0.026 & 0.24$\h 2.6  \cdot 10^{-4}$ \\
\hline
\end{tabular}
\ec
\end{table}

\end{document}